%% file: FKM2026_draft.tex
\documentclass[12pt]{article}
\PassOptionsToPackage{dvipsnames}{xcolor}
\usepackage{amsfonts, amsmath, amssymb, amsthm, appendix, array, bbm, bm, color, enumitem, forloop, framed, graphicx, lscape, mathrsfs, pdfpages, rotating, sectsty, setspace, threeparttable, tocloft,subfig,multirow,verbatim,rotating}

\usepackage{hyperref}
\hypersetup{pdfborder = {0 0 0},colorlinks=true,linkcolor=blue,citecolor=blue}
\usepackage[T1]{fontenc}
\usepackage{microtype}

\usepackage{tikz}
\usetikzlibrary{matrix,chains,positioning,decorations.pathreplacing,arrows}

\usepackage[round]{natbib}  

\usepackage[none]{hyphenat}

\theoremstyle{definition}
\newtheorem{remarkTmp}{Remark}
\newenvironment{remark}
	{\medskip  \begin{remarkTmp} 	}
	{

		\qed
		\end{remarkTmp}
	}

\newtheorem{egTmp}{Example}

\allowdisplaybreaks[1]

\setenumerate[1]{label=\bf(\arabic*)}
\setenumerate[2]{label=\bf(\roman*)}

	\DeclareMathOperator*{\argmax}{arg\,max}

	\newcommand{\E}{\mathbb{E}}

	\newcommand{\One}{\mathbbm{1}}

    \def\ddefloop#1{\ifx\ddefloop#1\else\ddef{#1}\expandafter\ddefloop\fi}
    \def\ddef#1{\expandafter\def\csname c#1\endcsname{\ensuremath{\mathcal{#1}}}}
    \ddefloop ABCDEFGHIJKLMNOPQRSTUVWXYZ\ddefloop
    \def\ddef#1{\expandafter\def\csname s#1\endcsname{\ensuremath{\mathsf{#1}}}}
    \ddefloop ABCDEFGHIJKLMNOPQRSTUVWXYZ\ddefloop
	\def\ddef#1{\expandafter\def\csname b#1\endcsname{\ensuremath{\bm{#1}}}}
	\ddefloop aAbBcCgGhHjJMpPqQsStTwWxXzZyY\ddefloop

\usepackage{booktabs}

\definecolor{darkgreen}{rgb}{0,0.5,0}

\begin{document}

\title{
	Robust A/B Decisions\thanks{
    The authors would like to thank Lars Hansen, Thomas Wiemann, and seminar participants at Chicago Booth and Meta for constructive feedback. We are also appreciative of our partner firm for making data and resources available for us to validate our ideas.      
    }
}

\author{
	Max H. Farrell\thanks{Department of Economics, UC Santa Barbara} \and Malika Korganbekova\thanks{Graduate School of Business, Stanford University} \and
	Sanjog Misra\thanks{Booth School of Business, University of Chicago}
}

\date{\today}

\maketitle

\bigskip

\begin{abstract}
A/B tests are standard in firm decision making. In the standard pipeline, experimental data is converted to a deployment decision by applying a $t$-test of the difference in means (the lift) and deploying the treatment if lift is positive and statistically significant. This common workflow answers the wrong question. We argue that firms need a decision rule for economic payoffs in the future deployment environment, not a test of equality in the experimental sample. We develop an ambiguity-averse decision framework in which each arm is evaluated by its ambiguity-penalized value over distributions close to the experimental outcome distribution. The resulting rule has a simple closed form thanks to the Donsker-Varadhan representation and it requires only the outcome data from a standard A/B test plus one interpretable parameter governing trust in the experiment. Our rule is thus no more difficult to implement than a $t$-test. A mean-variance approximation shows how the rule penalizes variability, while a connection to utility maximization shows it to be a certainty equivalent. We are able to perform a real-world evaluation of our proposed rule in the context of digital marketing using an archive of 552 advertising experiments from an anonymous US-based online platform. The proposed rule substantially reduces regret relative to conventional hypothesis testing. The results show that economically conservative, distribution-aware deployment rules can outperform statistical-significance rules in digital experimentation.
\end{abstract}


\setcounter{page}{0}
\thispagestyle{empty}

\newpage
\onehalfspacing
\section{Introduction}
	\label{sec:intro}

A/B tests are ubiquitous in industry. They are a workhorse of decision-making in digital firms. Major platforms run thousands each year to evaluate features, creatives, pricing, and recommendation policies \citep{kohavi2009online, bakshy2014designing, kohavi2020trustworthy, gordon2019comparison}. The analysis and decision pipeline is quite standardized: compute the difference in means, conduct a two-sided $t$-test, and deploy the treatment if the lift is positive and the $p$-value is below $\alpha = 0.05$. The procedure has the appeal of a clean and simple workflow. However, the $t$-test as a statistical tool is designed for a different question, not a deployment decision.

A $t$-test asks the scientific question of whether the experimental data contains enough evidence to conclude that the treatment and control have different means. The firm cares about the economic value of a policy in the deployment environment, that is, the environment in which the policy will actually be deployed at some future date. Both qualifiers do important work. ``Economic value'' insists that the evaluation criterion be the firm's payoff, not a feature of a sampling distribution. ``Deployment environment'' insists that the relevant evaluation be constructed in the context of the world the firm will actually face, not the world in which the experiment happened to be run. The conventional rule respects neither qualifier, and the two oversights compound.

For the first qualifier, consider a stylized example: take an incumbent policy $A$ with mean payoff $\$1$M and standard deviation $\$2$M and a candidate policy $B$ that delivers $\$1.2$M with certainty, with $n_A=n_B=100$ observations in each arm. In this case, the estimated lift is $\$200$K, the $t$-statistic is $1$ with a $p$-value of $0.32$. The conventional A/B testing pipeline keeps $A$. However, $B$ has a higher mean and zero risk, so any expected payoff maximizer would prefer it. This shows that the $t$-test based rule fails to consider the economics of the decision: the test correctly concludes that there is insufficient evidence to conclude that the two means are different, but this is the right answer to the wrong question. 

The second qualifier is more subtle. Deployment is not the experiment repeated at scale. An experiment runs in a particular window, on a particular customer mix, under particular competitive conditions. Deployment happens later, possibly on a different population, after novelty fades, or competitors respond. Randomization makes the experimental comparison internally credible. It does not make the experimental \emph{environment} the same as the deployment environment, and so it does not settle the question that matters to the firm. The experimental ranking of arms is informative about the deployment ranking only to the extent that the two environments are similar, and that similarity is itself uncertain.

Once this is taken seriously, the role of the experimental data changes. The conventional pipeline uses only the estimated means and variances, but the A/B test delivers an empirical \emph{distribution} of outcomes under each policy. This is informative about how dangerous it would be for the deployment environment to differ from the experimental one. A treatment whose gains are broad-based and stable is well-positioned against modest distributional drift: even if the deployment world tilts somewhat unfavorably, the bulk of the payoff is preserved. A treatment whose lift is driven by a small number of unusually favorable realizations is far more fragile, because the same tilt can shift mass into the unfavorable part of the distribution and reverse the policy ranking. Means summarize experimental lift; higher moments summarize deployment fragility. 

This reasoning suggests the form the decision problem and deployment rule should take. Making this precise, and providing an implementable workflow, is the goal of this paper. We model the firm as an ambiguity-averse decision maker. Ideally the firm would rank actions by their expected payoff in the deployment environment, but, acknowledging that the deployment environment is not pinned down by the experiment, the firm treats the experiment as a benchmark and assumes the deployment environment is similar. Formally, for each $a \in \{A,B\}$, the firm collects samples $Y_{a,i}$, $i=1, \ldots, n_a$, summarized by the experimental empirical distribution $P_a$. The deployment distribution, the unknown distribution of outcomes that would result from deploying arm $a$, is only assumed to lie in a neighborhood of $P_a$. The neighborhood is determined by a single parameter choice, $\lambda \geq 0$, whose interpretation is managerial: $\lambda$ records how much the firm trusts the experiment as a guide to deployment. When the firm believes the experiment is a tight guide to deployment it should compare expected payoffs at the experimental distribution itself, and when the firm believes deployment could differ more it should evaluate each arm under a more pessimistic neighborhood. Operationalizing our deployment rule requires the choice of $\lambda$ in the same way that the conventional workflow requires a threshold for the $p$-value. Section \ref{sec:lambda} gives a full discussion of $\lambda$.

The formal decision framework, detailed in the next section, yields a simple, closed-form expression for the value of each arm $a \in \{A,B\}$, by applying the Donsker--Varadhan variational identity \citep{donsker1983asymptotic}. The decision rule is to deploy the arm with the higher ``robust'' (or penalized) value $V_\lambda(a)$, where
\begin{equation}
	\label{eqn:value}
    V_\lambda(a) \;=\; -\frac{1}{\lambda}\, \log\!\left( \frac{1}{n_a} \sum_{i=1}^{n_a} e^{-\lambda Y_{a,i}} \right).
\end{equation}
Implementing this decision rule is no more onerous than the conventional $t$-test rule: nothing beyond the raw outcome data already collected on a standard A/B testing dashboard is required: no prior, no structural model, no auxiliary covariates, and only a single parameter chosen by the decision-maker.

The distribution that attains \eqref{eqn:value} is an exponential tilt of the empirical distribution, $dQ_{a,\lambda}^{\mathrm{wc}} \propto e^{-\lambda Y}\, dP_a$, which shows exactly how the experimental data directly informs the worst-case scenario the firm is willing to entertain. Note that this worst-case distribution is specific to each arm, so differences between the entire distributions for $A$ and $B$ inform the firm's decision. For example, in the stylized experiment above, $B$ has zero variance in the experiment, so the worst-case distribution is the experimental distribution, perfectly capturing that the payoff from $B$ is certain.

The neighborhood is not selected by an exogenous adversary, but rather is determined via the experimental payoff distribution being stress-tested in the direction the firm cares about. At $\lambda = 0$, the rule reduces to the comparison of sample means. As $\lambda$ grows, the robust value $V_\lambda(a)$ smoothly grows more conservative, allowing for larger deviations from the experimental distribution. This is a conservative decision criterion, but it is conservative in an economic sense rather than a purely statistical one. It asks what could go wrong under plausible misspecification of the deployment environment, and it chooses the policy whose payoff is best after accounting for that possibility.

The robust value \eqref{eqn:value} can also be written as a certainty equivalent. This shows that the same decision rule can arise from familiar risk-averse utility maximization: under constant absolute risk aversion and a known payoff distribution, the planner would rank policies using the same functional form. Our motivation is different, but the link is useful as it formalizes that volatile, skewed, or downside-exposed policies are  precisely the policies whose values are most sensitive to such distribution misspecification. Ambiguity aversion can be reinterpreted as a particular form of risk adjustment in this way.

We conduct a large-scale empirical evaluation of our proposed decision rule using an archive of 552 advertising A/B tests from an anonymous US-based online platform.\footnote{\textbf{Disclaimer:} The Data Use Agreement with the anonymous platform allows the platform to ask to re-scale the numbers for data anonymity purposes. However, the platform does not have the right to impact the results or the research questions the academics work on.} We compare our rule to the standard practice of hypothesis testing as well as several others, including the empirical success rule, which is recovered by \eqref{eqn:value} with $\lambda=0$. We use both random and geographical sample splitting to create held out testing data, ensuring that comparisons do not suffer from overfitting, and compare each rule to the ex-post optimal decision in that sample using average regret. We find that conventional $t$-testing deploys $B$ in fewer than two percent of experiments and yields regret comparable to random decisions that ignore the data completely. Under our rule regret improves by roughly 25\%, yielding massive improvements in decision making and revenue.

This paper connects to several different strands of literature. The first is the literature on the design and use of A/B tests in digital markets, which has received considerable attention from researchers across industry \citep{kohavi2009online, kohavi2020trustworthy, gordon2019comparison, gordon2023close}, statistics \citep{Goldberg-Johndrow2017_WP}, economics \citep{Azevedo-Deng-Olea-Weyl2019_AERPP}, and marketing \citep{Feit-Berman2019_MS, Joo-Chiong2026_MS}. Our paper contributes to this literature by showing how to incorporate ambiguity about the deployment environment into the decision framework and how to evaluate decisions based on their welfare consequences rather than by the operating characteristics of statistical tests. 

The latter point is the focus of a large literature criticizing the use of hypothesis testing for decision making. See \cite{Manski2011_ARE,Manski2019_AmerStat,manski2019patient} for excellent discussions and further references, including work on decisions under ambiguity. This criticism is well understood, and we do not claim novelty in arguing that statistical significance is not itself an appropriate decision criterion. Particularly closely related is \citet{Feit-Berman2019_MS}, who reframe A/B testing as a profit-maximization problem and derive an optimal ``test-and-roll'' design that explicitly trades off the opportunity cost of experimentation against the expected value of deploying the superior treatment. Their analysis provides an important precedent for replacing hypothesis-testing conventions with economically grounded experimental decisions. Our focus is complementary: taking an experiment of a given size as completed, we study the deployment decision when the experimental outcome distribution may be an imperfect guide to the future deployment environment. We account explicitly for this distributional shift and obtain a tractable rule that uses the full outcome distribution of each arm. Other approaches in the decision literature include minimax-regret rules \citep{Stoye2012_JoE,Tetenov2016_WP,Manski2019_AmerStat} and Bayesian and empirical-Bayes decision rules \citep{Goldberg-Johndrow2017_WP,Azevedo-Mao-Olea-Velez2023_JET}. Except in the limiting case $\lambda=0$, in which our rule reduces to the empirical-success rule studied by \citet{Stoye2012_JoE}, our distributionally robust deployment rule has not, to our knowledge, been studied in this literature. 

Finally, we use ideas from the literatures on robust decisions \citep{hansen2008robustness} and distributionally robust optimization \citep{Kuhn-Shafiee-Wiesemann2025_ActaNum}, both of which are motivated by ambiguity aversion. We adapt this machinery to A/B testing and demonstrate its value in deployment decisions in digital marketing. Although we restrict attention to A/B tests, which are binary decisions, our ideas can be applied to multi-valued treatments without change and could be extended to other settings, including continuous treatments, bandit problems, exploiting heterogeneity or targeting, and other areas where statistical significance should not be used as a basis for policy decisions. 

The remainder of the paper proceeds as follows. Section \ref{sec:framework} develops the decision framework and derives the value \eqref{eqn:value}, including discussion of the choice of $\lambda$. Section \ref{sec:empirical} shows the empirical results from the online platform archive. Section \ref{sec:conclusion} concludes.

\section{A/B Decision Framework}
	\label{sec:framework}

This section formalizes the A/B testing problem from first principles as an economic decision under ambiguity and derives the decision rule described in Section \ref{sec:intro}. We show that the closed form solution \eqref{eqn:value} is obtained as the worst-case that explicitly formalizes the conservative approach to accounting for the ambiguity over the deployment distribution.

\subsection{Decision Problem Set Up}

The idealized version of the firm's goal is to choose an arm/action $a^\star \in \{A,B\}$ to maximize the expected payoff after deployment at scale ($A$ and $B$ are also referred to as control and treatment, incumbent and challenger, or simply denoted $\{0,1\}$). Each policy $a \in \{A,B\}$ induces a distribution $Q^{\mathrm{dep}}_a$ over scalar, individual-level per-period payoffs $Y_a$ that the firm will realize after deployment. Ideally, the firm would select the arm that maximizes the expected payoff according to the true \emph{deployment distribution} $Q^{\mathrm{dep}}_a$. Thus the firm would like to select $a^\star$ according to 
\begin{equation}
	\label{eqn:ideal problem}
    a^\star = \argmax_{a \in \{A,B\} } \E_{Q^{\mathrm{dep}}_a} \left[ Y_a \right].
\end{equation}
(Throughout $Y_a$ can be replaced by $u(Y_a)$ for a utility function $u(\cdot)$. See also \eqref{eqn:cara} below.) 

However this problem is not feasible because $Q^{\mathrm{dep}}_a$ is unknown. To learn about $Q^{\mathrm{dep}}_a$, $a \in \{A,B\}$, the firm conducts a randomized experiment that yields data on each arm: $\{Y_{a,i}\}$, $i=1, \ldots, n_a$, which collectively define the \emph{experimental distribution} $P_a$. The firm then uses the data $\{Y_{a,i}\}$ to learn about the distribution of payoffs under each arm, and importantly in our case, uses the entire $P_a$ to learn about the deployment distribution. Throughout, $P_a$ is the empirical distribution: $P_a = \sum_{i=1}^{n_a} \delta_{Y_{a,i}} / n_a$, where $\delta_y$ is a point mass at $y$. All quantities derived from each $P_a$ are also sample quantities, and therefore computable with the data at hand. See Remark \ref{rem:continuity} for discussion.

We specifically model this learning to respect the ambiguity about the deployment distribution. We posit that the firm, or manager, is willing to assume that $Q^{\mathrm{dep}}_a$ is, in a divergence sense, ``close'' to $P_a$ and adopts a maximin approach to the decision problem. The firm has a multiplier preference that measures the firm's ambiguity tolerance. Specifically, we assume that the firm evaluates each arm according to the entropy-penalized robust value and the decision problem is therefore 
\begin{equation}
	\label{eqn:robust problem}
	a^\star = \argmax_{a \in \{A,B\} } \min_{Q \ll P_a} \left\{ \E_Q \left[Y_a\right] + \frac{1}{\lambda} D_{\mathrm{KL}}(Q \| P_a) \right\},
\end{equation}
where $D_{\mathrm{KL}}(Q \| P) = \int \log(dQ / dP) dQ$ is the Kullback-Leibler divergence (Remark \ref{rem:continuity}). The parameter $\lambda$ determines the planner's willingness to trade off the expected payoff against relative-entropy departures from the reference model. The choice of $\lambda > 0$ is the key parameter of our framework and governs the ambiguity tolerance. Alternative distributions are penalized by relative entropy at rate $1/\lambda$. Small $\lambda$ corresponds to a large penalty, so that the firm essentially trusts the experiment as a guide to deployment, whereas for large $\lambda$ the firm entertains large drifts and is highly conservative. Interpreting and choosing $\lambda$ are discussed in Section \ref{sec:lambda} below and further in the empirical evaluation. Note that in this formulation, the same $\lambda$ is applied to each arm, which is natural because the firm is concerned about the deployment environment and not the experimental environment. The decision maker applies the same shadow price ($1/\lambda$) to model misspecification for each available action. This formulation is familiar from distributionally robust optimization \citep{Kuhn-Shafiee-Wiesemann2025_ActaNum} and robust decisions \citep{hansen2008robustness}, but applied to a new context. In the conventional Hansen–Sargent parameterization, $1/\lambda$ is the relative-entropy multiplier.

\begin{remark}[Connection to $t$-testing]
    \label{rem:ttesting ambiguity}
	To connect our idea to the usual A/B testing framework, we note that the $t$-test considers a very different, and narrow, sense of ambiguity: the distribution $Q^{\mathrm{dep}}_a$ is assumed to be the population distribution from which $P_a$ is a sample. In this setting, $Q^{\mathrm{dep}}_a$ is also modeled as unknown and different from $P_a$, but only in this repeated sampling sense. Note that as $n_a \to \infty$ this ambiguity vanishes completely, whereas we maintain that $Q^{\mathrm{dep}}_a$ can still be different from $P_a$. See Remark \ref{rem:asymptotics} for the role of asymptotic theory in our work and Remark \ref{rem:precision} for incorporation of sample size. 
\end{remark}

\begin{remark}[KL Divergence and Absolute Continuity]
    \label{rem:continuity}
    Modeling all distributions as empirical is most faithful to the decision problem facing the firm of using current experimental data to learn about future payoffs, both of which are finite sample settings (e.g., conducting an experiment this month and measuring revenue in the quarter after deployment). However, for empirical distributions $Q$ and $P$ generally $D_{\mathrm{KL}}(Q \| P_a)$ is infinite as the support points do not agree, meaning \eqref{eqn:robust problem} is technically restricted to a reweighting of the observed data. Our goal is to arrive at the closed form solution \eqref{eqn:value} and \eqref{eqn:our rule}, which is well-defined for empirical distributions, and so we do not dwell on this technicality. For all discussions of implementation all distributions, moments, and values, are computed from the sample. For example, $\mu_a = \E_{P_a}[Y_a] = \sum_{i=1}^{n_a} Y_{a,i}/{n_a}$, $a \in \{A,B\}$. Continuity formalities can be addressed by first working in the population and then appealing to asymptotic theory, as discussed in Remark \ref{rem:asymptotics}.
\end{remark}

\begin{remark}[Ambiguity Set Formulation]
	\label{rem:ambiguity set}
	Throughout we take \eqref{eqn:robust problem} to be the primitive formulation of the decision problem. However, it is useful to connect this to the optimization over an ambiguity set. For a fixed $\lambda$, the inner minimization in \eqref{eqn:robust problem} shares a minimizer with the constrained optimization Kullback-Leibler ball around $P_a$:
	\begin{equation}
		\label{eqn:constrained problem}
		\min_{Q}  \Big\{ \E_Q \left[Y_a\right] : Q \ll P_a, D_{\mathrm{KL}}(Q \| P_a) \leq \eta_a \Big\},
	\end{equation}
	where the radius $\eta_a \geq 0$ is determined by $\lambda$ and $P_a$. In our formulation, $\lambda$ is fixed and common because it is the firm's preference parameter, while the implied $\eta_a$ is necessarily arm-specific. 
\end{remark}

\subsection{A Feasible Decision Rule}

The key to operationalizing the decision rule is that the variational problem in \eqref{eqn:robust problem}, the inner minimization, admits a closed-form solution by applying the Donsker-Varadhan representation of the Kullback-Leibler divergence \citep{donsker1983asymptotic}:
\begin{equation}
	\label{eqn:dv}
    \min_{Q \ll P_a} \left\{ \E_Q\!\left[Y_a\right] + \frac{1}{\lambda} D_{\mathrm{KL}}(Q \,\|\, P_a) \right\} = -\frac{1}{\lambda} \log \left( \frac{1}{n_a} \sum_{i=1}^{n_a} e^{-\lambda Y_{a,i}} \right) =: V_\lambda(a),
\end{equation}
which is exactly the definition of $V_\lambda(a)$ initially stated in \eqref{eqn:value}. The ambiguity-averse planner therefore ranks policies by the right-hand side of \eqref{eqn:dv}, and thus, recalling that $P_a$ is the empirical distribution in the experiment, the decision rule is given by 
\begin{equation}
	\label{eqn:our rule}
	a^\star = \argmax_{a \in \{A,B\} } V_\lambda(a).
\end{equation}
The multiplier value $V_\lambda(a)$, previewed in the Introduction, is easily computed for each arm of a randomized experiment. Just as with the usual testing-based framework, only the raw data and a single preference parameter is required from the decision-maker. We do not use a structural model or a prior.

Given a choice of $\lambda$ (see Section \ref{sec:lambda}), the decision rule \eqref{eqn:our rule} is well-defined and fully feasible. The rule is also interpretable. The worst-case distribution $Q^{\mathrm{wc}}_{a,\lambda}$ that attains the minimum in \eqref{eqn:robust problem} is an exponential tilt of the empirical distribution, $dQ^{\mathrm{wc}}_{a,\lambda} \propto e^{-\lambda Y}\, dP_a$, which shows exactly how the experimental data directly informs the worst-case scenario the firm is willing to entertain. Note that this worst-case distribution is specific to each arm, so differences between the entire distributions for $A$ and $B$ inform the firm's decision. For example, in the stylized experiment above, $B$ has zero variance in the experiment, and this will imply a worst-case distribution extremely close to the experiment, perfectly capturing that the payoff from $B$ is certain.

However, to better understand $V_\lambda(a)$ and the parameter $\lambda$, and therefore the associated decision rule, it is useful to make two connections. First, recognizing that $V_\lambda(a)$ contains the cumulant generating function, for small values of $\lambda$ a Taylor approximation yields
\begin{equation}
	\label{eqn:taylor}
    V_\lambda(a)  =  \mu_a  -  \frac{\lambda}{2} \sigma_a^2   +   O(\lambda^2),
\end{equation}
where $\mu_a = \sum_{i=1}^{n_a} Y_{a,i}/{n_a}$ and $\sigma_a^2 = \sum_{i=1}^{n_a} (Y_{a,i} - \mu_a)^2/{n_a}$ are the experimental mean and variance. Note that this mean-variance expansion is exact (in the population) for Gaussian data, but large sample Gaussian approximations (of the sampling distribution of the mean) do not remove the higher order terms: decision robustness is about the changing of the distribution under deployment, not statistical uncertainty about a population distribution. See Remarks \ref{rem:asymptotics} and \ref{rem:precision}.

Truncating to the leading terms, the form of \eqref{eqn:taylor} is intuitive: a higher (sample) mean increases the valuation of that arm, but higher variance (for a fixed $\lambda$) indicates that $Q^{\mathrm{dep}}_a$ may deviate further from $P_a$, and thus, because we take a worst-case approach in \eqref{eqn:constrained problem}, this lowers the valuation of the arm. Intuitively, a higher variance can be indicative of deployment fragility for a variety of reasons, but a leading example is effect heterogeneity. If all customers in the experiment respond similarly to the treatment (low variance), then the firm can expect similar payoff in deployment. If the treatment is highly effective for some customers and ineffective for others (high variance), then customer mix changes in deployment will change the ranking of the arms.

This approximation also helps with interpretation of $\lambda$ as being measured relative to the mean and variance, and is thus case-specific: the units must match the decision problem. If $1/\lambda$ is interpreted as a shadow price, we see that the price must be measured in the corresponding units and will require the firm to select $\lambda$ for a given decision, not globally.

Retaining only the mean and variance yields a compact approximation to the decision rule \eqref{eqn:our rule}. Conveniently, this approximation uses only summary statistics and can therefore be applied whenever the raw data is not available, such as reevaluation of past decisions. This is useful for calibration of $\lambda$, as we discuss below. In our evaluation in Section \ref{sec:empirical}, we find that this mean-variance approximation makes quite different decisions than \eqref{eqn:our rule}, indicating that higher moments are decision-relevant in our data, relative to the scale of $\lambda$, but yields similar performance in terms of overall regret.

The second connection that is useful for intuition is to classical utility-maximization decision theory under risk: ambiguity aversion can be shown to be equivalent to conventional risk aversion. Our view of the firm is as a risk-neutral decision-maker, i.e., only the mean matters in \eqref{eqn:ideal problem}, and we arrive at the decision rule \eqref{eqn:our rule} by modeling ambiguity about the deployment distribution. However, the same decision rule arises if we model the firm as an expected-utility maximizer with constant absolute risk aversion (CARA) and Bernoulli utility
\begin{equation}
    \label{eqn:cara}
    u(y) = -e^{-\lambda y}/\lambda.
\end{equation}
In this case the certainty equivalent of the gamble $Y_a \sim P_a$ is \emph{exactly} $V_\lambda(a)$ of \eqref{eqn:value} and \eqref{eqn:our rule}. The penalty parameter $\lambda$ in our framework is then mathematically \emph{exactly} the Arrow-Pratt coefficient of absolute risk aversion in this expected-utility formulation. This connection is useful for interpretation, by connecting to familiar economic machinery, and moreover, can be used to calibrate $\lambda$ based on elicited or estimated risk preferences.

It is important to remember that although the ultimate decision rules are identical, the interpretations are conceptually distinct, as is the meaning of $\lambda$. In the CARA utility theory, ``risk'' is in the conventional sense of an aversion to randomness of outcomes, whereas in our framework the planner is worried about distributional shift. A risk-neutral planner can be uncertain about the post-deployment distribution just as a risk-averse planner could fully trust the experimental distribution as a guide for the future. The final rules are the same, but the setup of \eqref{eqn:cara} does \emph{not} follow from replacing $Y_a$ with $u(Y_a)$ in \eqref{eqn:ideal problem} or \eqref{eqn:constrained problem}. In fact, doing so would yield \emph{both} classical risk aversion and ambiguity regarding the distribution, which may be of use in other application areas: combining the two is straightforward and coherent.

\begin{remark}[Connection to $t$-testing]
The expansion \eqref{eqn:taylor} is useful for connecting our decision to the usual $t$-test, because the $O(\lambda^2)$ terms of \eqref{eqn:taylor} depend on the higher order cumulants, and so the mean-variance approximation is exact for Gaussian data, which is also a standard assumption for $t$-testing. Both methods then require only the sample means and variances as inputs for computation, but crucially, (i) comparing $V_\lambda(A)$ vs. $V_\lambda(B)$ uses the \emph{difference} of the variances, whereas the $t$-test uses the \emph{sum}, as in Section \ref{sec:rules}, (ii) $V_\lambda(a)$ uses the \emph{variance} of the outcome, whereas the $t$-test uses the \emph{standard error} of the estimate of the mean. See also Remark \ref{rem:precision}.
\end{remark}

\begin{remark}[Asymptotic Theory]
	\label{rem:asymptotics}
	The decision rule \eqref{eqn:our rule} is stated (and implemented) using only sample data, matching our motivation of studying the A/B testing problem. From a statistical theoretic perspective, however, the reliance entirely on sample quantities leads to some technical issues, such as in Remark \ref{rem:continuity}. These can easily be overcome formally by starting entirely from population quantities and then using asymptotic theory to justify the use of sample quantities. One would take $Q^{\mathrm{dep}}_a$ in \eqref{eqn:ideal problem} to be an unknown/future population distribution and assume that the experimental distribution $P_a$ is the empirical distribution of $n_a$ i.i.d. draws from a population distribution $P_a^\star$ such that $Y_a \sim P_a^\star$ is a continuous random variable with $\E[|Y_a|] < \infty$ and $\E_{P_a^\star}[\exp\{-\lambda Y_a\}] < \infty$. We then replace $P_a$ by $P_a^\star$ in \eqref{eqn:robust problem}, so that this becomes the population version of the decision problem, and apply the Donsker-Varadhan representation. This yields that decisions are to be made by comparing $V_\lambda^\star(a) = -(1/\lambda) \log \left( \E_{P_a^\star} \left[e^{-\lambda Y_a}\right]\right)$. Finally, because $V_\lambda^\star(a)$ is unknown, we replace it by the sample analogue $V_\lambda(a)$ of \eqref{eqn:value} and appeal to the law of large numbers, which yields $V_\lambda(a) \to_P V_\lambda^\star(a)$. Under further moment conditions we also can quantify the sampling uncertainty of this estimate. This statistical theory is standard, and arrives at the same feasible decision rule, and so we do not dwell on the details. It is worth noting that this is straightforward for each experiment because we assume i.i.d. data, but extending this to the empirical evaluation in Section \ref{sec:empirical} is more complicated because we have many experiments, each with its own distribution and unknown, complex dependencies between experiments due to subject overlap and spatial and temporal dependence. 
\end{remark}

\subsection{The Decision Rules We Study}
	\label{sec:rules}
	
Here we give a short outline of the decision rules we study for concreteness and reference, collecting our proposal and conventional methods. These will be tested empirically in the sequel. To emphasize their similarities and differences we unify the notation as much as possible. Recall that for each arm $a \in \{A,B\}$, $\mu_a$, $\sigma_a^2$, and $n_a$ are the \emph{sample} mean, variance, and size, respectively, and that $V_\lambda(a)$ is computed as in \eqref{eqn:our rule}. To clarify each decision rule's dependence on a tuning parameter, let $d(\cdot)$ be a binary decision function where $d = 1$ corresponds to deploying arm $B$ and $d=0$ represents maintaining the status quo arm $A$. To match standard practice in industry, only if $B$ exhibits a higher experimental mean is deployment entertained. We implement this throughout, even though it distorts the rule above somewhat.

\begin{itemize}
	\item {\bf Robust decision.} Implementing our proposed rule \eqref{eqn:our rule} yields the decision function
	\[
		d_{\textsc{ce}}(\lambda) = \One\Big\{ \mu_B - \mu_A  > 0 \Big\}  \cdot \One \Big\{ V_\lambda(B) - V_\lambda(A) > 0 \Big\}.
	\]
	We denote this $d_{\textsc{ce}}$ due to the connection to the certainty equivalent under CARA utility. 
	
	\item {\bf Mean-variance approximation.} Using the approximation of \eqref{eqn:taylor} truncated to the two leading terms yields the rule
	\[
		d_{\textsc{mv}}(\lambda) = \One\Big\{ \mu_B - \mu_A  > 0 \Big\}  \cdot \One\Big\{ \mu_B - \mu_A  -  \frac{\lambda}{2} \big(\sigma_B^2 - \sigma_A^2\big) > 0 \Big\}.
	\]
	This rule coincides with $d_{\textsc{ce}}(\lambda)$ for Gaussian data. Note that $d_{\textsc{mv}}(\infty)$ corresponds to deploying $B$ only if it strictly dominates $A$: higher return and lower variance.
	
	\item {\bf Empirical success.} Selecting the arm with the higher mean is the simplest and most aggressive deployment rule we consider:
	\[
		d_{\textsc{es}} = \One\Big\{ \mu_B - \mu_A  > 0 \Big\} 
	\]
	The threshold of zero could be replaced by some other data-independent value, but in practice this is not done. This rule is also the limit with no ambiguity aversion: $d_{\textsc{es}} = d_{\textsc{mv}}(0) = d_{\textsc{ce}}(0)$.
	
	\item {\bf Hypothesis testing.} For a significance level $\alpha$ and corresponding critical value $c_\alpha$, the rule is
	\[
		d_{\textsc{pv}}(\alpha) = \One\Big\{ \mu_B - \mu_A  -  c_\alpha \big(\sigma_B^2/n_B + \sigma_A^2/n_A\big)^{1/2} > 0 \Big\}.
	\]
	We set $\alpha = 0.05$, and hence $c_\alpha = 1.96$, throughout our evaluation.\footnote{Given the standard practice in industry of only considering $B$ if it exhibits a higher experimental mean this is a one-sided testing problem, but it is also standard to use two-sided critical values and report two-sided $p$-values.} Note that $d_{\textsc{pv}}(1) = d_{\textsc{es}} = d_{\textsc{mv}}(0) = d_{\textsc{ce}}(0)$ and the same holds for any $\alpha$ as $\min\{n_A,n_B\} \to \infty$ if the population means are unequal. For positive $\lambda$, it is otherwise impossible to equate $d_{\textsc{pv}}(\alpha)$ with $d_{\textsc{ce}}(\lambda)$ or $d_{\textsc{mv}}(\lambda)$, because $d_{\textsc{pv}}(\alpha)$ depends on the sum of the variances whereas $d_{\textsc{ce}}(\lambda)$ or $d_{\textsc{mv}}(\lambda)$ use the difference.
	
	\item {\bf Randomized decision.} For comparison, we also include a purely random decision which ignores the data completely. Let $U$ be a coin flip that returns each of $A$ and $B$ with probability 1/2 and set 
	\[
		d_{\textsc{rn}} = \One\Big\{ U = B \Big\}  .
	\]
	This serves as a baseline: any data-driven rule should yield higher payoffs.	
\end{itemize}

\begin{remark}[Statistical Precision]
    \label{rem:precision}
    Our decision rule does not take into account statistical precision which can be unappealing. For a simple example, consider the empirical welfare maximization rule ($\lambda = 0$) which simply selects the higher mean. It is counterintuitive that a decision maker would be indifferent between two experiments, one with 10 observations and one with 10,000, if the means are identical. It is natural for decision makers to value precise estimates in some way, but it is not immediately obvious how to incorporate this into a formal decision framework. Recent work by \cite{Chernozhukov-Lee-Rosen-Sun2025_WP} proposes one approach to this problem which adds a penalty to the decision rule based on the standard error of the mean, essentially incorporating a preference for precision into the decision rule. Our work is complementary: we focus on the distributional shift problem, and do not incorporate statistical precision into the decision rule. However, it would be natural to combine these approaches, which amounts to adding a statistical penalty to $V_\lambda(a)$. This would require a second preference parameter, but one with a distinct interpretation: whereas $\lambda$ encodes the planner's concern for distributional shift, the new parameter, say $\gamma$, would encode the planner's concern for statistical precision. Recall from Remark \ref{rem:asymptotics} that $V_\lambda(a)$ is asymptotically Gaussian under standard conditions. Let $S_a$ be the standard error of $V_\lambda(a)$. Then the decision rule \eqref{eqn:our rule} is replaced by
	\[
		a^\star = \argmax_{a \in \{A,B\} } \big\{ V_\lambda(a) - \gamma S_a \big\}.
	\]
	We will not pursue this here, and defer formal analysis to \cite{Chernozhukov-Lee-Rosen-Sun2025_WP}. It is important to note that, as made clear in \eqref{eqn:taylor}, when $n_a \to \infty$, $S_a \to 0$, and so the statistical penalty vanishes, but the distributional shift penalty does not. This is a key distinction between the two concerns.
\end{remark}

\begin{remark}[Risk Aversion]
    With the decision rules defined, it is useful to return to the discussion of risk aversion. Our proposed rule, $d_{\textsc{ce}}(\lambda)$ is motivated by the ``risk'' of a changing deployment environment, but as discussed above at \eqref{eqn:cara}, is identical to what a decision maker with CARA utility would implement, for risk aversion parameter $\lambda$, but remember that the meaning of $\lambda$ is different. The $t$-test guards against the ``risk'' of type I error, which should not be conflated with risk aversion in any economically meaningful sense. The decision $d_{\textsc{pv}}(\alpha)$ depends on (i) standard errors instead of variances and (ii) the sum instead of the difference. Utility maximizers do not have preferences over lift directly, i.e. decisions are not made based upon $\E[u(Y_B - Y_A)]$, but rather they care about the lift in expected utility, $\E[u(Y_B)] - E[u(Y_A)]$. Only for a risk- and ambiguity-neutral firm are these the same, in which case the variances are irrelevant anyway. A risk-neutral planner would use $\lambda = c_\alpha = 0$ and follow the empirical success rule to be optimal \citep{Stoye2012_JoE}. As we show in the empirical results, $d_{\textsc{pv}}(\alpha)$ can yield risk-\emph{loving} decisions (see the simple example in the Introduction).
\end{remark}

\begin{remark}[Other Decision Rules]
    As aforementioned, there is a large literature criticizing statistical testing for decision making and proposing other decision rules. We focus on the set above because our goal is to compare our rule to the commonly used $t$-test approach. Other works use minimax theory \citep{Manski2011_ARE,Manski2019_AmerStat}, Bayesian methods \citep{Goldberg-Johndrow2017_WP,Feit-Berman2019_MS,Azevedo-Mao-Olea-Velez2023_JET}, or empirical Bayes \citep{Azevedo-Deng-Olea-Weyl2019_AERPP}, though we caution that some take the experiment as providing a signal about the lift directly, $(\mu_B - \mu_A)$, which means that only the sum of the variances can be considered, leading to $t$-statistics being used, instead of properly studying each arm separately. A long line of work in economics uses minimax regret to study optimal decision rules \citep{Stoye2012_JoE,Tetenov2016_WP,Manski2019_AmerStat,Joo-Chiong2026_MS}.
\end{remark}

\subsection{Choosing $\lambda$}
	\label{sec:lambda}

Choice of $\lambda$ is the key input required by the decision maker to operationalize the rule $d_{\textsc{ce}}(\lambda)$ of \eqref{eqn:our rule} based on $V_\lambda(a)$, or the approximation $d_{\textsc{mv}}(\lambda)$. This is not a tuning parameter in the statistical sense, and there is no notion of ``optimality'' in $\lambda$. Rather, this parameter must encode the preferences of the planner: $\lambda$ reflects the degree of the planner's trust in the experiment as a guide to deployment payoffs. By the definition of the problem, $\lambda$ is an economic preference parameter, and is therefore difficult to scale, since utility itself need not have a universal scale. Decision-makers must recognize that $\lambda$ is specific to a firm or decision problem, as it depends on $P_a$ and $P_b$, the experimental context. 

It is helpful to use the mean-variance approximation of \eqref{eqn:taylor}: $V_\lambda(a) \approx \mu_a - \lambda \sigma_a^2 / 2$. In this form, we see that $\lambda/2$ represents the relative weight the planner gives to variance, i.e., the rate at which the planner is willing to trade off expected revenue versus variability. In this case, because all other features of the distribution are ignored (or the data is Gaussian), variance directly indicates potential for $Q^{\mathrm{dep}}_a$ to differ from $P_a$. For example, if the data are Gaussian and $\sigma_a^2=0$, then it must be that $Q^{\mathrm{dep}}_a$ and $P_a$ coincide, and thus so do their means, the ultimate object the decision maker cares about.

For interpretation, first consider the two extremes. If $\lambda = 0$, the value of each arm is given by the sample mean, $V_0(a) = \mu_a = \sum_{i=1}^{n_a} Y_{a,i} / n_a$, and thus our rule \eqref{eqn:our rule} also collapses to the empirical success rule. The intuition is clear: when $\lambda = 0$, the experiment is thought to be a perfect guide to deployment, so $Q^{\mathrm{dep}}_a = P_a$, and then \eqref{eqn:ideal problem} can be solved by selecting the higher experimental mean. This is the most aggressive deployment strategy, and as above, is minimax optimal under further assumptions. On the other hand, as $\lambda \to \infty$, the planner allows arbitrarily large drift and $V_\lambda(a) \to \min \{Y_{a,i} \}$ and the decision rule becomes ultra pessimistic, corresponding to classical Manski bounds: ranking policies by their worst possible outcome (assuming the sample minimum equals a known population support bound). Here $d_{\textsc{mv}}(\infty)$ corresponds to selecting $B$ only when it has both a higher return and lower variance. In general, $\lambda = \infty$ can be too pessimistic for practical use but may give a sense of scale for $\lambda$.

Several approaches can be used in practice to select a $\lambda$, depending on the availability both of past data and of information about the planner. First, absent any information of any kind, the decision can be evaluated across a grid of $\lambda$ and the sensitivity reported. This is also useful practice in evaluating past decisions, which can guide the choice of grid. 

Second, planners who already have a stated tolerance for shift can use this directly. This preference could also be elicited, and an interesting direction to extend this paper would be to conduct a conjoint study to estimate $\lambda$ at different firms or for different populations of decision makers. One can also link to risk aversion using \eqref{eqn:cara}, as discussed there, which is useful for elicitation and calibration. 

Third, if past experimental data is available, one can use (features of) the distribution of \emph{implied} $\lambda$. Past experimental decisions can then be used to empirically calibrate how the firm has traded off lift and variability, whether interpreted as ambiguity about the distribution or classical risk aversion. 

Fourth, requiring both past experimental and follow-up data, the firm can compute the distance $D_{\mathrm{KL}}(Q^{\mathrm{dep}}_{m,a^\star_m} \| P_{m,a^\star_m})$, where $m$ indexes the past data and $a^\star_m$ is the chosen action in that past setting.\footnote{This distance can be computed by binning or kernel smoothing, for example, to avoid the discreteness as discussed in Remark \ref{rem:continuity}.} These distances can then be used to compute a value for $\eta$ in \eqref{eqn:constrained problem}, such as the average or worst-case, and this can be directly translated to $\lambda$ by mapping the solutions of the two optimization problems under further assumptions.

In our empirical analysis below we combine the first and third approaches. Using the control arms from our experimental data archive, under the notion that the firm has long implemented this status quo to reflect its preferences, we calibrate $\lambda$ so that the firm's decisions are on average consistent with the common heuristic that the risk premium is roughly 10\% of the value.\footnote{This form of moment matching is widely used in the
literature. For example, \citet{lettau2008declining} calibrate relative risk
aversion to match the mean equity premium together with other aggregate
asset-pricing moments; \citet{collindufresne2017asset} choose risk aversion so
that the model-implied equity Sharpe ratio is comparable to its empirical
counterpart; and \citet{lochstoer2022volatility} set the risk-aversion parameter
directly by matching the equity premium. Related work also calibrates risk
aversion to generate an empirically plausible equity premium under alternative
preference specifications \citep{driessen2025picapm}.} The risk premium, as implied by \eqref{eqn:taylor}, is $\lambda \sigma^2_A/2$, and as a fraction of the mean we should have that $10\% \approx (\lambda \sigma^2_A/2) / \mu_A$. Solving this, we compute, across all our experiments, $\widehat\lambda_m = 0.1 (2 \mu_{A,m})/\sigma^2_{A,m}$, where $m$ indexes the experiments. We then take the average of the $\widehat\lambda_m$ and obtain $\widehat\lambda = 0.000208$, yielding a rough approximation to the firm's preferences in this data. We repeat this using five percent ($\widehat\lambda = 0.000104$) and twenty-five percent ($\widehat\lambda = 0.000521$), to explore sensitivity. As a reminder, the goal here is to obtain several $\lambda$ values for this particular empirical evaluation, not to provide a definitive or optimal $\lambda$ for future use.

\begin{remark}[Connection to $t$-testing]
Setting the $p$-value threshold of 0.05 is so ingrained in practice that one often forgets that this is also a tuning parameter and is chosen in advance based on the planner's preferences. The key difference is that the choice of $p$-value reflects a desire for statistical conservativeness, in the form of type I error, and has no economic content. This also means that it does not vary according to context or application, as type I error is always the same concept. Conventional wisdom selects $p = 0.05$, but there is no a priori reason why this choice is sensible for decision making. Indeed, some have argued that in digital marketing decisions a much smaller value should be used \citep{Azevedo-Deng-Olea-Weyl2019_AERPP}. From a decision-theoretic view there is no reason to favor type I over type II error in this way, and in any case, the economic value of these errors should be considered. In one-sided testing, a $p$-value of 0.5 (i.e. 50\%) yields regret optimal decisions with Gaussian data \citep{Stoye2012_JoE,Tetenov2016_WP} as this collapses the decision to the empirical success rule.
\end{remark}

\section{Empirical Analysis}
    \label{sec:empirical}

\subsection{Data}
	\label{sec:data}

To evaluate our decision rule, and compare to the others in Section \ref{sec:rules}, we use an archive of 552 advertising A/B tests conducted by an anonymous US-based online platform. Each experiment randomly assigns customers to treatments $A$ or $B$ and records a per-customer monetary outcome.\footnote{We exclude A/A tests (system validation), Do No Harm guardrail experiments, and multi-arm comparisons (though our method applies here as well). Some tests evaluate multiple variants against a single control.} Table \ref{tab:summary_552} summarizes the data: Panel A reports design features, B shows summary statistics for the outcome, and C tabulates relevant $p$-values. The sample sizes in the experiments are large but the effects are small relative to the noise. The median experiment involves 3,516,000 customers, but the median within-arm standard deviation is 280 against a median outcome mean of \$35. Lifts cluster tightly around zero (median $-0.03$, IQR $[-0.28,\,0.15]$). Because of this, conventional statistical significance thresholds are rarely achieved: only 7.6\% of the 552 comparisons are significant at the 5\% level. A firm using $d_{\textsc{pv}}(0.05)$ would deploy $B$ in fewer than one in ten of its experiments, regardless of the economic value at stake. This pattern is typical for digital experiments where small per-person effect sizes are economically significant due to scale \citep{Taddy-etal2016_JBES,Azevedo-Deng-Olea-Weyl2019_AERPP}.

\begin{table}[htbp!]
\caption{Summary Statistics}\label{tab:summary_552}
\centering
\begin{threeparttable}
\small
\input{inputs/table1_552}
\begin{tablenotes}[flushleft]
\footnotesize
\item \textit{Notes.} Panel~A: sample sizes (thousands). Panel~B: per-customer outcomes. Panel~C: two-sided $t$-test $p$-values.
\end{tablenotes}
\end{threeparttable}
\end{table}

\begin{figure}[htbp!]
\caption{Experimental Lift vs.\ Variance Difference}\label{fig:scatter_552}
\centering
\begin{minipage}{0.85\textwidth}
\centering
\includegraphics[width=\linewidth]{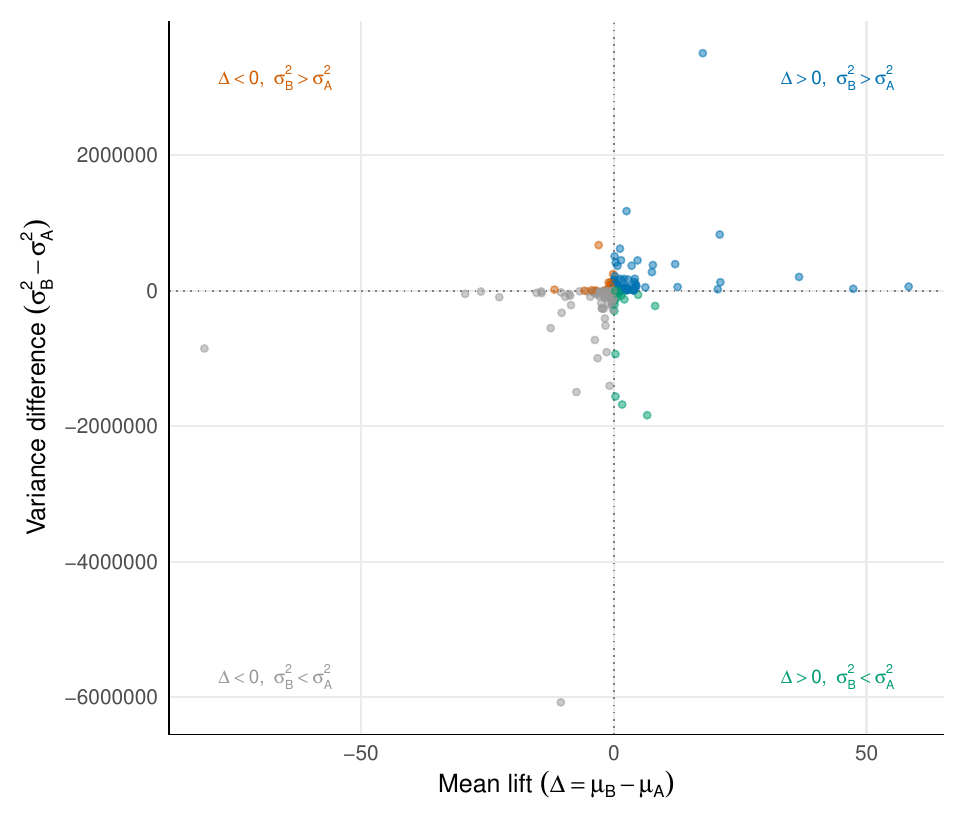}
{\smallskip \footnotesize \textit{Notes.} Each point is one of 552 experiments. Horizontal axis: $\mu_B - \mu_A$. Vertical axis: $\sigma_B^2 - \sigma_A^2$. Colors correspond to quadrants.\par}
\end{minipage}
\end{figure}

Figure \ref{fig:scatter_552} shows each experiment in our data in the $(\mu_B - \mu_A,\,\sigma_B^2 - \sigma_A^2)$ plane. Experiments populate all four quadrants, showing that the way the different rules of Section \ref{sec:rules} treat means and variances has large consequences. In real world practice, firms only deploy $B$ if the lift is positive, as encoded in the rules above, which restricts attention to quadrants I and IV for the analysis below.

\subsection{Evaluation Methodology}
	\label{sec:method}

Our evaluation of the decision rules relies on two key ingredients: (i) sample splitting for a fair comparison and (ii) regret for a concrete measure of quality. Sample splitting is used to ensure that decision rules are judged based on realized payoffs from customers the decision-maker did not see. Fundamentally, this is a prediction problem, where the decision-maker uses the experimental data $\{P_A, P_B\}$ to predict if arm $A$ or $B$ yields higher mean revenue under the deployment distribution. Like any prediction problem, the proper benchmark is out of sample validation. Otherwise, overfitting would render the conclusions unreliable. We implement sample splitting in two different ways: randomly and geographically.

Random splitting serves as a benchmark and sanity check. When each experiment is randomly split into training and testing, the two samples are drawn from the same distribution, and so the training data should be a good guide to (the proxy for) deployment: $Q^{\mathrm{dep}}_a$ should be close to $P_a$ for each arm. This is the sample analogue of the idealized problem of \cite{Stoye2012_JoE}, where the empirical success rule is minimax optimal. We operationalize this using cross fitting. Each of the 552 experiments is partitioned randomly into five (roughly) equal subsamples. In each of five rounds, four folds are used for training and one for testing and all results are then averaged across all five folds (i.e. cross fitting). In the training data we evaluate all the rules of Section \ref{sec:rules}, computing $V_\lambda(a)$, means, variances, $p$-values, etc, and record each decision in the form of $d \in \{A,B\}$. Following convention, we use $\alpha=0.05$ for testing-based decisions. For our rules $d_{\textsc{ce}}(\lambda)$ and $d_{\textsc{mv}}(\lambda)$ we take calibrated $\lambda$ as in Section \ref{sec:lambda}, including zero, so the grid we use is $\lambda = \{0, 0.000104, 0.000208, 0.000521\}$. The fifth fold serves as the test data where we evaluate which arm has the higher average revenue and compare this to the decision for each rule. Results are then averaged over the five folds.

Geographical splitting is more realistic and better reflects the problem at hand, where the experimental and deployment settings may differ. Here we divide each experiment into training and testing by state.\footnote{The data comes from the 50 US states as well as the District of Columbia, Puerto Rico, U.S. Virgin Islands, and Guam, and thus has 54 jurisdictions. We use the word ``state'' throughout for simplicity.} First, one state is held out as the test data, while the other 53 are the training data. Just as above, the rules are fit on the training data and then tested in the held out state. Finally, results are averaged across all states. We also do this in reverse, where the single state is the training data and the remaining 53 serve as the test data. The latter reflects the practice of performing a small-scale experiment in one environment and then deploying the chosen arm to the entire customer base. In either case, forward or reverse, the training and testing data are drawn from different distributions, giving very clear motivation to our approach. The geographical splitting also allows us to explore how the rules perform when the training and testing distributions differ in a way that is interpretable: customers in one state may be qualitatively different than those in another state.

Finally, we measure decision quality using \emph{regret}, which is standard in the literature. Regret measures the quality of the decisions not just by their correctness (in identifying the better arm) but by using the economic value of the decision. Regret captures the gap between the payoff the oracle achieves and the payoff the rule delivers. Here the oracle observes the noisy measure of the truth in the test data: $\Delta^{\textrm{test}} := \mu_B^\mathrm{test} - \mu_A^\mathrm{test}$ and deploys $B$ if and only if the lift is positive in the test data. For a decision rule $d$, regret is
\begin{equation*}
    \max(\Delta^{\textrm{test}}, 0) - \Delta^{\textrm{test}} \cdot d.
\end{equation*}
A false positive (deploying a treatment that hurts) generates regret equal to the magnitude of the harm; a false negative (withholding a treatment that helps) generates regret equal to the benefit foregone. Unlike classification accuracy, regret weights each mistake by its economic size, so deploying a treatment that costs \$2 per customer is penalized more heavily than deploying one that costs \$0.01. Regret is always averaged to reflect the same customer population, matching the firm-wide deployment decision. State-level results are weighted by state population prior to averaging for this reason.

\subsection{Regret Analysis}

Table \ref{tab:regret_552_nl} shows the main regret results, using random cross fitting. We show that both $d_{\textsc{ce}}(\lambda)$ and its approximation $d_{\textsc{mv}}(\lambda)$ substantially outperform hypothesis testing in terms of regret. In this data, hypothesis testing performs only slightly better than purely random choice (the exact values for $d_{\textsc{rn}}$ vary with the seed for the random number generator; the table reports the average across ten seeds). Hypothesis testing studies the wrong question, so it is not surprising that it does poorly. This is particularly true given the salient features of digital marketing, i.e., very small lifts with heavy tails. Here our results reinforce prior work criticizing testing for decision-making, as discussed above.

Turning to our decision rules, we see that $d_{\textsc{ce}}(\lambda)$ and $d_{\textsc{mv}}(\lambda)$ perform similarly to each other and across $\lambda$. The empirical success rule ($d_{\textsc{es}} = d_{\textsc{ce}}(0) = d_{\textsc{mv}}(0)$) has the lowest regret. Because random sample splitting would not be expected to produce large differences between the training (experiment) sample and testing (deployment) sample, this finding is in line with \cite{Stoye2012_JoE}. Higher $\lambda$ prunes marginal deployment decisions, but in all cases our decision rule yields significant revenue improvements over testing.

\begin{table}[htbp]
\caption{Mean Regret -- Random Sample Splitting}\label{tab:regret_552_nl}
\centering
\begin{threeparttable}
\input{inputs/cw_cv5_per_experiment_regret_new_lambdas.tex}
\begin{tablenotes}[flushleft]
\footnotesize
\item \textit{Notes.} Per-experiment customer-level average regret averaged across five-fold random cross fitting. 
\end{tablenotes}
\end{threeparttable}
\end{table}

Results for the geography-based cross fitting are shown in Table \ref{tab:geo_552_nl} and provide direct evidence for our proposed methodology. Here the ambiguity over the payoff distribution is directly interpretable: the customers in one state may be qualitatively different than others. This is borne out in the findings: regret is not increasing with $\lambda$ and $d_{\textsc{es}}$ is not the best rule. This shows directly how it is important for firms to consider the difference between the experimental and deployment distributions. To highlight the interesting findings following the sanity check in the random splitting case, Table \ref{tab:geo_552_nl} shows results for $d_{\textsc{ce}}(\lambda)$ and $d_{\textsc{pv}}(0.05)$.

\begin{table}[htbp]
\caption{Mean Regret -- Geographic Sample Splitting}\label{tab:geo_552_nl}
\centering
\begin{threeparttable}\centering
\input{inputs/geo_ef_combined_552_new_lambdas}\begin{tablenotes}[flushleft]
\footnotesize
\item \textit{Notes.} Forward: train on national sample excluding focal state, test on focal state. Reverse: train on focal state, test on national sample excluding focal state. States weighted by customers within each experiment, then experiments averaged equally. This produces the difference in the scale of regret.
\end{tablenotes}
\end{threeparttable}
\end{table}

\begin{remark}[Economic Magnitude]
	\label{rem:magnitude}
	The preceding results measure the improvement in average out-of-sample regret per customer. The natural scale for a stand-alone deployment is the firm's active-customer base. This firm has approximately 20 million customers, and so, because the reduction in regret relative to conventional hypothesis testing is roughly \$0.20--\$0.54 per customer (from Tables \ref{tab:regret_552_nl} and \ref{tab:geo_552_nl}), the revenue-equivalent gain would be $20\text{ million}\times[\$0.20,\$0.54]  =  [\$4\text{ million},\$10.8\text{ million}]$. However, aggregating this figure across experiments is difficult: customers may be in multiple experiments, not every experiment applies to every customer because of targeting to categories or locations, treatments may affect overlapping components of the customer experience, later interventions may substitute for or supersede earlier ones, and treatment effects may not persist indefinitely. None of the relevant factors are identifiable in our data, and indeed, are difficult to find in most contexts. Naively assuming full additivity across all experiments would imply a firm-wide gain of $[\$2.2\text{ billion},\$5.5\text{ billion}]$, which is not plausible. However, if only 5\% of this gain is realized, this still represents approximately \$110--\$275 million, illustrating the substantial economic consequences of testing-based deployment decisions.
\end{remark}

\subsection{Where the Decision Rules Differ}
    \label{sec:decision-differences}

To understand the source of the regret improvement, we compare the decisions themselves using the random-splitting cross fitting analysis. Table \ref{tab:2x2_552_nl} compares our proposed rule $d_{\textsc{ce}}(\lambda)$ to hypothesis testing purely in terms of deployment decisions. First, we see that $d_{\textsc{pv}}(0.05)$ almost never deploys $B$, although when it does, the decision is always correct in the test data. Though as noted above, it is incorrect to label this failure of $d_{\textsc{pv}}(0.05)$ as being overly ``conservative'' or ``risk-averse'' in the usual sense. It is the missed opportunities of $d_{\textsc{pv}}(0.05) = 0$ that are so costly. For example, at $\lambda=0$, $d_{\textsc{ce}}(0) = 1$ for 232 experiments, and nearly three quarters of these are the correct decision in the test data. In the same vein, when $d_{\textsc{ce}}(\lambda) = d_{\textsc{pv}}(0.05) = 0$ (the final column), the test data agrees in the majority of cases. We omit $d_{\textsc{mv}}(\lambda)$ for brevity, but this rule makes different decisions than both $d_{\textsc{ce}}(\lambda)$ and $d_{\textsc{pv}}(0.05)$.

\begin{table}[htbp]
\caption{Decision Decomposition: $d_{\textsc{ce}}(\lambda)$ vs.\ $d_{\textsc{pv}}(0.05)$}\label{tab:2x2_552_nl}
\centering
\begin{threeparttable}
\centering
\scalebox{0.88}{%
\input{inputs/cw_cv5_2x2_ce_pv_per_experiment_new_lambdas--MAX.tex}}
\begin{tablenotes}
\footnotesize
\item \textit{Notes.} Cell counts of experiments with parentheses showing the share of these with positive lift in the test data, $\mu_B^\mathrm{test} - \mu_A^\mathrm{test} > 0$. Rules are marked as $d=1$ if this is true for a majority of five folds.
\end{tablenotes}
\end{threeparttable}
\end{table}

\begin{figure}[htbp!]
\caption{Deployment Decisions: $d_{\textsc{ce}}(\lambda)$ vs.\ $d_{\textsc{mv}}(\lambda)$ at $\lambda = 0.000104$}\label{fig:mv_deploy_552}
\centering
\begin{minipage}{0.85\textwidth}
\centering
\includegraphics[width=\linewidth]{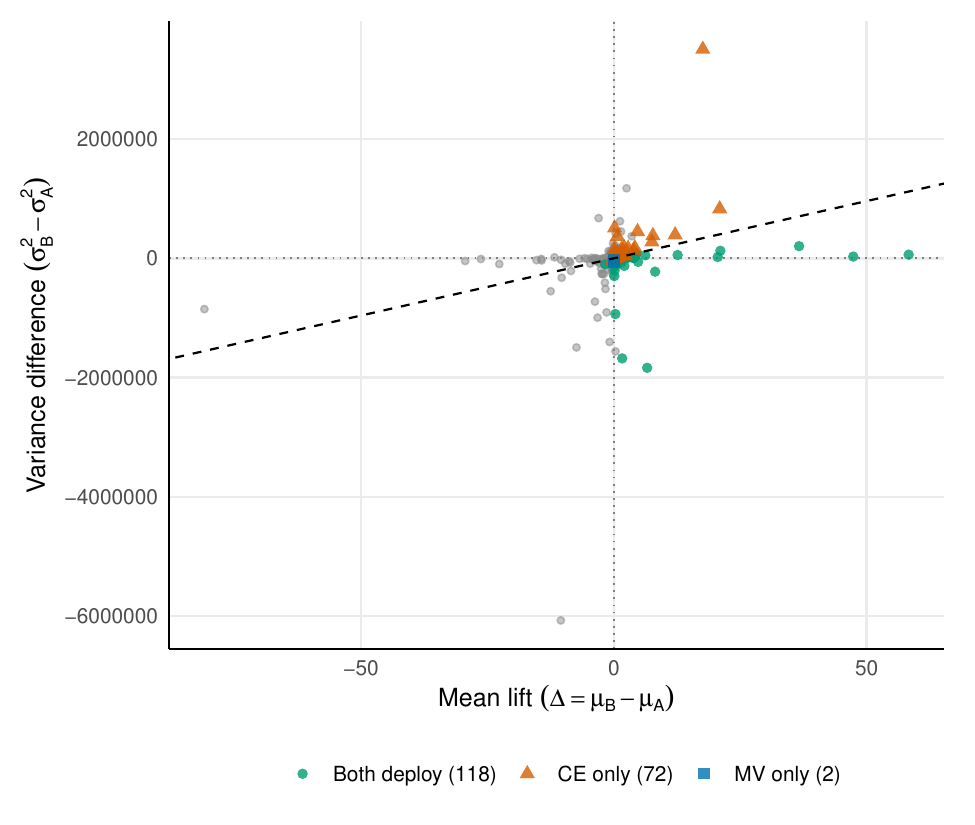}
{\smallskip \footnotesize \textit{Notes.} Green: both $d_{\textsc{ce}}(\lambda)$ and $d_{\textsc{mv}}(\lambda)$ set $d=1$ (118). Orange triangles: $d_{\textsc{ce}}(\lambda)=1$ only (72). Blue squares: $d_{\textsc{mv}}(\lambda)=1$ only (2). Grey: $d=0$ for both (360). Dashed line: $d_{\textsc{mv}}(\lambda)$ boundary $\sigma_B^2 - \sigma_A^2 = (2/\lambda)\Delta$ with slope $2/\lambda = 19{,}200$. \par}
\end{minipage}
\end{figure}

Figure \ref{fig:mv_deploy_552} shows the same scatter plot as Figure \ref{fig:scatter_552} but labeled according to decisions. Both $d_{\textsc{mv}}(\lambda)$ and $d_{\textsc{es}}$ have simple decision regions in this plane. The empirical success rule, $d_{\textsc{es}} = d_{\textsc{mv}}(0) = d_{\textsc{ce}}(0)$, always selects the higher mean, and so $d_{\textsc{es}}=1$ for quadrants I and IV: the boundary is the vertical line at $\mu_B - \mu_A = 0$. In comparison, $d_{\textsc{mv}}(\lambda) =1 $ if $(\mu_B - \mu_A) > \lambda (\sigma_B^2 - \sigma_A^2)/2$, corresponding to all experiments below a ray through the origin with slope $2/\lambda$, including \emph{some} of quadrant I (where $\sigma_B^2$ is not so large as to outweigh the positive lift) and \emph{all} of quadrant IV, for which arm $B$ dominates $A$. The full decision rule $d_{\textsc{ce}}(\lambda)$ does not have a simple geometric boundary, but Figure \ref{fig:mv_deploy_552} shows that $d_{\textsc{ce}}(\lambda)=1$ for many experiments above the boundary line for $d_{\textsc{mv}}(\lambda)$, highlighting that the rules do not agree, although they yield comparable regret performance as reported in Table \ref{tab:regret_552_nl}.

These results show that the higher order terms neglected in \eqref{eqn:taylor} are important in this application area. This is for two reasons. First, despite the large sample sizes, the underlying data are not Gaussian. Most customers spend little or nothing, while a small number spend heavily. This is typical for digital marketing applications. Second, the scale of $\lambda$ is relative, and important. Even though $\lambda = 0.000104$ may appear ``small'', it is not relative to the size of the variance. For example, in one experiment the lift is \$21.22, which is quite large, but $(\sigma^2_B - \sigma^2_A) \approx 1,016,000$, and so the variance penalty for $d_{\textsc{mv}}(0.000104)$ is $(0.000104/2) \times 1{,}016{,}000 = \$52.83$, which is 2.49 times the lift, so $d_{\textsc{mv}}(0.000104)=0$ in this case, while $d_{\textsc{ce}}(0.000104) = 1$ by using the entire distribution. 

\begin{figure}[htbp!]
\caption{Experiments Deployed by $d_{\textsc{pv}}(0.05)$}\label{fig:pv_deploy_552_nl}
\centering
\begin{minipage}{0.85\textwidth}
\centering
\includegraphics[width=\linewidth]{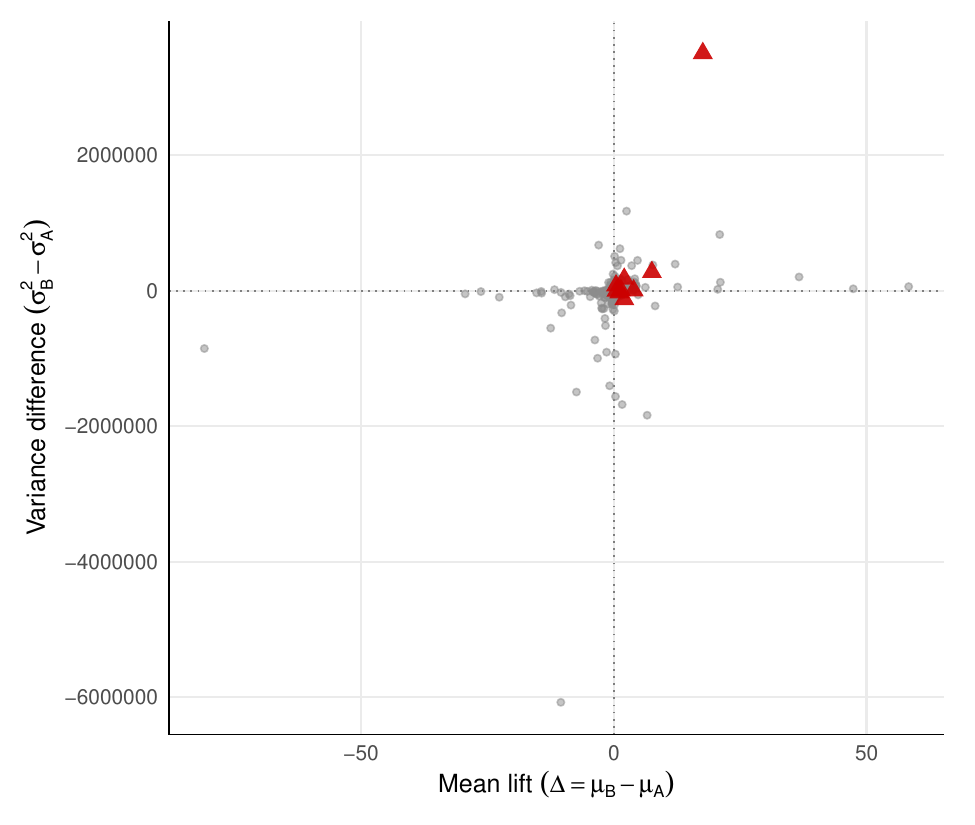}
{\smallskip \footnotesize \textit{Notes.} Red triangles: 9 experiments with $d_{\textsc{pv}}(0.05)=1$ (positive lift and $p < 0.05$). Grey: $d_{\textsc{pv}}(0.05)=0$ (543). The p-value rule sets $d=1$ for fewer than 2\% of experiments. \par}
\end{minipage}
\end{figure}

Finally, for comparison Figure \ref{fig:pv_deploy_552_nl} shows the deployment decisions for $d_{\textsc{pv}}(0.05)$, which also does not have a simple geometric boundary. First, we see that very few experiments yield $d_{\textsc{pv}}(0.05)=1$, as above. Further, these are not those with highest lift: $d_{\textsc{pv}}(0.05)$ may ignore high-return arms even with low risk/variance if the \emph{sum} of the variances is large, and in this way can even exhibit decisions consistent with risk-loving preferences. 

\section{Conclusion}
    \label{sec:conclusion}

This paper reframes A/B testing as an economic deployment decision under ambiguity rather than as a hypothesis-testing exercise. The central problem is that the experimental distribution is only a guide to the future deployment distribution, and statistical significance does not measure either the economic value of an arm or its fragility to distributional shift. We propose a simple robust decision rule that ranks each arm by its penalized payoff. The rule admits a closed-form certainty-equivalent representation, uses only the data already collected in a standard A/B test, and requires a single parameter that governs how much the decision maker trusts the experiment as a guide to deployment. The rule is no more difficult to implement than the standard practice of hypothesis testing. 

In the archive of 552 advertising experiments, the conventional testing rule deploys very rarely and produces regret comparable to ignoring the data entirely. By contrast, the robust rule and its mean-variance approximation substantially reduce regret across random and geographic validation designs. These findings suggest that many digital experiments contain economically useful information that is discarded when decisions are filtered through statistical significance thresholds. The broader lesson is that firms should replace tests with decision rules that encode payoffs, uncertainty about deployment, and tolerance for fragility. 

Future work can extend this framework to other areas of experimental design and analysis. One can consider more complex experimental settings, such as measuring spillovers or network effects, or capturing treatment effect heterogeneity for personalization. Extensions to noncompliance would be relevant in many business and economic settings. Just as we suggest replacing hypothesis testing at the analysis stage, one could also consider implications for design and replace power calculations. Empirically, it would be useful in applications to conduct surveys or conjoints to measure the ambiguity aversion of real-world decision makers and establish norms around decision making in industry.

\section{References}
\singlespacing
\begingroup
\renewcommand{\section}[2]{}	
\bibliography{FKM--Bibliography}{}
\bibliographystyle{ecta}
\endgroup

\end{document}

%% file: inputs/table1_552.tex
\begin{tabular}{lrrrrr}
\toprule
\multicolumn{6}{l}{\textit{Panel A: Experimental Design}} \\
\midrule
 & Mean & SD & Q25 & Median & Q75 \\
\midrule
  Treatment arm ($n_B$, thousands) & 4,508 & 7,022 & 232 & 1,765 & 6,622 \\
  Control arm ($n_A$, thousands) & 4,621 & 7,606 & 232 & 1,758 & 6,621 \\
  Total per experiment (thousands) & 9,129 & 14,386 & 467 & 3,516 & 13,243 \\
\midrule
\multicolumn{6}{l}{\textit{Panel B: Outcomes}} \\
\midrule
  Treatment mean ($\bar{Y}_B$) & 61.87 & 74.58 & 18.71 & 35.69 & 63.25 \\
  Control mean ($\bar{Y}_A$) & 61.96 & 74.96 & 18.70 & 36.01 & 62.97 \\
  Mean lift ($\Delta$) & -0.09 & 6.06 & -0.28 & -0.03 & 0.15 \\
  Percent lift (\%) & -0.45 & 9.17 & -0.82 & -0.15 & 0.44 \\
  Treatment SD ($\sigma_B$) & 352.2 & 275.7 & 181.5 & 280.7 & 418.5 \\
  Control SD ($\sigma_A$) & 362.9 & 302.0 & 184.0 & 282.5 & 421.4 \\
\midrule
\multicolumn{6}{l}{\textit{Panel C: Statistical Significance}} \\
\midrule
 & \multicolumn{2}{c}{Count} & \multicolumn{3}{c}{Share (\%)} \\
\cmidrule(lr){2-3} \cmidrule(lr){4-6}
  $p < 0.01$ & \multicolumn{2}{c}{16} & \multicolumn{3}{c}{2.9} \\
  $0.01 \leq p < 0.05$ & \multicolumn{2}{c}{26} & \multicolumn{3}{c}{4.7} \\
  $0.05 \leq p < 0.10$ & \multicolumn{2}{c}{30} & \multicolumn{3}{c}{5.4} \\
  $p \geq 0.10$ & \multicolumn{2}{c}{480} & \multicolumn{3}{c}{87.0} \\
\bottomrule
\end{tabular}

%% file: inputs/cw_cv5_per_experiment_regret_new_lambdas.tex
\begin{tabular}{lrrrrr}
\toprule
 & & \multicolumn{4}{c}{Mean Regret} \\
\cmidrule(lr){3-6}
$\lambda$ &  & $d_{\textsc{ce}}(\lambda)$ & $d_{\textsc{mv}}(\lambda)$ & $d_{\textsc{pv}}(0.05)$ & $d_{\textsc{rn}}$ \\
\midrule
  0 &  & 1.086 & 1.086 & 1.486 & 1.512 \\
  0.000104 &  & 1.117 & 1.154 & 1.486 & 1.512 \\
  0.000208 &  & 1.132 & 1.170 & 1.486 & 1.512 \\
  0.000521 &  & 1.139 & 1.225 & 1.486 & 1.512 \\
\bottomrule
\end{tabular}

%% file: inputs/geo_ef_combined_552_new_lambdas.tex
\begin{tabular}{lrrrr}
\toprule
 & \multicolumn{2}{c}{Forward} & \multicolumn{2}{c}{Reverse} \\
 & \multicolumn{2}{c}{\footnotesize (train national, test state)} & \multicolumn{2}{c}{\footnotesize (train state, test national)} \\
\cmidrule(lr){2-3} \cmidrule(lr){4-5}
$\lambda$ & $d_{\textsc{ce}}(\lambda)$ & $d_{\textsc{pv}}(0.05)$ & $d_{\textsc{ce}}(\lambda)$ & $d_{\textsc{pv}}(0.05)$ \\
\midrule
  0 & 4.153 & 4.695 & 1.052 & 1.265 \\
  0.000104 & 4.173 & 4.695 & 1.047 & 1.265 \\
  0.000208 & 4.150 & 4.695 & 1.054 & 1.265 \\
  0.000521 & 4.177 & 4.695 & 1.053 & 1.265 \\
\bottomrule
\end{tabular}

%% file: inputs/cw_cv5_2x2_ce_pv_per_experiment_new_lambdas--MAX.tex
\begin{tabular}{lrrrr}
\toprule
 & \multicolumn{2}{c}{$d_{\textsc{ce}}(\lambda) = 1$} & \multicolumn{2}{c}{$d_{\textsc{ce}}(\lambda) = 0$} \\
\cmidrule(lr){2-3} \cmidrule(lr){4-5}
$\lambda$ & $d_{\textsc{pv}}(0.05)=1$ & $d_{\textsc{pv}}(0.05)=0$ & $d_{\textsc{pv}}(0.05)=1$ & $d_{\textsc{pv}}(0.05)=0$ \\
\midrule
  0 & 9 (100.0\%) & 223 (73.5\%) & 0 (0.0\%) & 320 (17.8\%) \\
  0.000104 & 9 (100.0\%) & 181 (79.0\%) & 0 (0.0\%) & 362 (21.5\%) \\
  0.000208 & 9 (100.0\%) & 179 (78.8\%) & 0 (0.0\%) & 364 (22.0\%) \\
  0.000521 & 9 (100.0\%) & 170 (80.6\%) & 0 (0.0\%) & 373 (22.5\%) \\
\bottomrule
\end{tabular}